\documentclass[twocolumn,nofootinbib, aps,groupedaddress]{revtex4-1}

\usepackage{amsmath, amssymb, graphicx, color, comment, cancel}
\usepackage[colorlinks=true,linkcolor=blue,citecolor=blue]{hyperref}

\definecolor{purple}{rgb}{0.5 ,0, 0.7}
\definecolor{bluegreen}{rgb}{0, 0.45, 0.35}

\begin{document}

\begin{flushright}
IPMU 17-0093
\end{flushright}
\title{Oscillating Affleck-Dine condensate and its cosmological implications}


\author{Fuminori Hasegawa}
\email[fuminori@icrr.u-tokyo.ac.jp]{}
\author{Masahiro Kawasaki}
\email[kawasaki@icrr.u-tokyo.ac.jp]{}
\affiliation{ICRR, The University of Tokyo, Kashiwa, 277-8582, Japan\\Kavli IPMU (WPI), TODIAS, The University of Tokyo, Kashiwa, 277-8583, Japan}


\date{\today}

\begin{abstract}
We study the dynamics of the Affleck-Dine field after inflation in more detail. After inflation, the Affleck-Dine field inevitably oscillates around the potential minimum. This oscillation is hard to decay and can cause accidental suppression of the consequential baryon asymmetry. This suppression is most effective for the model with non-renormalizable superpotential $W_\text{AD}\sim\Phi^4$ ($\Phi$: Affleck-Dine field). It is found that the Affleck-Dine leptogenesis in high-scale inflation, which suffers from serious gravitino overproduction, becomes workable owing to this effect.
\end{abstract}

\pacs{}

\maketitle



%


\section{Introduction}

The Big Bang nucleosynthesis (BBN) successfully explains the abundances of light elements in our universe if we adopt the baryon density determined by observations of the cosmic microwave background (CMB), which indicates that the baryon asymmetry of the universe is $n_{B}/s\sim10^{-10}$ at the beginning of the BBN. On the other hand, before the BBN, to give an explanation to the horizon problem, flatness problem and the origins of the tiny inhomogeneity of the universe, the accelerated expansion of the early universe, called inflation is considered to occur. However, inflation must wash out the pre-existing baryon asymmetry of the early universe. Therefore, we need a mechanism to generate the adequate baryon asymmetry after inflation and before the BBN.

There exist various types of mechanisms to generate the baryon number density in the early universe. In particular, Affleck-Dine baryo/leptogenesis \cite{AFFLECK1985361,DINE1996291} is a promising candidate in supersymmetric theory (SUSY) because it is realized in the minimal supersymmetric standard model (MSSM). In the MSSM, there are a lot of flat directions with a non-zero $B$--$L$ charge, called Affleck-Dine field (AD field). During and after inflation, the AD field could have a large vacuum expectation value (VEV) due to the negative Hubble induced mass. As the energy of the universe decreases in matter domination era after inflation, the soft SUSY breaking mass for the AD field overcomes the negative Hubble induced mass and the AD field starts to oscillate around the origin coherently. At the same time, the phase-dependent part of the potential (A-term) for the AD field becomes effective and ``kicks" the AD field into the phase direction. As a result the AD field rotates in the complex field plane. Since the $B$--$L$ number density is determined by the ``angular momentum" in the field plane, $B$--$L$ asymmetry of the universe is produced by this mechanism. Finally, the AD field decays into quarks, leptons and their anti-particles and generates $B$--$L$ asymmetry in thermal plasma \cite{1475-7516-2013-01-017}.  The $B$--$L$ asymmetry is further converted to the baryon asymmetry through the sphaleron effect\cite{FUKUGITA198645,KUZMIN198536}.   

To estimate the produced baryon asymmetry, we have to follow the dynamics of AD field in the cosmological background. Conventionally, it is assumed that AD field keeps following the time-dependent potential minimum adiabatically. Actually, however, the AD field inevitably oscillates around its vacuum. Its amplitude and period are characterized by the K\"ahler mixing between inflation sector and the AD field. We discuss the dynamic in detail and find that this oscillation causes an accidental suppression of $B$--$L$ number density depending on the model parameters. This effect is most efficient when the dynamics of the AD field $\Phi$ is governed by the non-renormalizable superpotential $W_\text{AD} \sim \Phi^n$ with $n=4$. In this paper, we apply this suppression mechanism to the minimal Affleck-Dine leptogenesis scenario~\cite{MURAYAMA1994349} where $LH_u$ flat direction is used ($n=4$). Consequently, we show that we can avoid the gravitino problem in the minimal Affleck-Dine leptogenesis after high-scale inflation by an appropriate choice of model parameters.

The remaining parts of this paper are as follows. First, in Sec.~2, we briefly review the Affleck-Dine baryogenesis and derive the conventional evaluation of $B$--$L$ number density. Next, we discuss the oscillation dynamics of AD field just after inflation and estimate its contribution to $B$--$L$ number density numerically, In Sec.~4, we apply this effect to the minimal Affleck-Dine leptogenesis scenario. Finally, we conclude in Sec.~5. 

\section{Affleck-Dine baryogenesis}

Let us review the conventional Affleck-Dine baryogenesis scenario. First, we discuss the scalar potential for the AD field. The AD field is exactly flat in renormalizable level unless SUSY is broken. However, non-renormalizable terms and SUSY breaking effect lift its potential. The non-renormalizable superpotential for the AD superfield $\Phi$ is cast as
\begin{align}
W_{\rm{AD}}=\lambda\frac{\Phi^n}{nM_p^{n-3}},
\end{align}
where $\lambda$ is a coupling constant, $n(\geq4)$ is a certain integer which is determined by specifying a flat direction. Here we take the Planck mass $M_p$ as the cutoff scale of the non-renormalizable terms. Then, the scalar potential for AD field including a soft SUSY breaking term and a Hubble induced mass term is given by
\begin{align}\label{pot}
V_{\rm{AD}}(\phi,\phi^*)&=(m_\phi^2-cH^2)|\phi|^2\\
&+\left(a_m\frac{\lambda m_{3/2}\phi^n}{nM_p^{n-3}}+\rm{h.c.}\right)+\frac{\lambda^2|\phi|^{2n-2}}{M_p^{2n-6}},
\end{align}
where   $\phi=\varphi e^{i\theta}$ is a scalar component of the AD superfield $\Phi$, $H$ is a Hubble parameter, $m_{3/2}$ is a gravitino mass and $m_{\phi}$ is a soft SUSY breaking mass for AD field, $a_m$, $c$ are $\mathcal{O}(1)$ parameters. In particular, the value of $c$ is responsible for the K\"ahler mixing between the AD superfield and the inflation sector. For example, we assume the inflation sector consists of two superfields $I$ and $S$, which are the inflation and so-called ``stabilizer", quartic K\"ahler mixing between the AD field, in general, is written as 
\begin{align}
K_{\rm mix}=\frac{c_1}{M_p^2}|\Phi|^2|I|^2+\frac{c_2}{M_p^2}|\Phi|^2|S|^2.
\end{align}
 Here $c_1,c_2$ are $\mathcal{O}(1)$ constants. In this case, the constant $c$ is evaluated in terms of $c_1,c_2$ as \cite{PhysRevD.74.063507}
\[
  c = \begin{cases}
    c_I\equiv3(c_2-1) & ({\rm during\ inflation}) \\
    c_M\equiv \frac{3}{2}\left(c_1+c_2-1\right)& ({\rm after\ Inflation})
  \end{cases}
\]
Note that the value of $c$ is different for during/after inflation in general.\footnote{If inflation is driven by single superfield with a K\"ahler mixing $c'|\Phi|^2|I|^2$, take $c_1=c_2=c'$.\label{f}} Hereafter we consider the case $c>0$. 

Let us follow the cosmological evolution of the AD field. At first, during inflation where $cH>m_\phi$, the AD field $\phi$ develops a large VEV due to the negative Hubble induced mass $\sim -cH^2$ as
 \begin{align}\label{b}
\varphi_0(t)|_{t<t_e}\simeq\left(\frac{\sqrt{c_I/(n-1)}}{\lambda}H_IM_p^{n-3}\right)^{\frac{1}{n-2}},
\end{align}
where $t_e$ is the time when inflation ends. 

After inflation, the energy density of the universe is dominated by the coherent oscillation of the inflaton and the Hubble parameter starts to decrease. Therefore, the minimum of the AD field is also time-dependent and approaches to zero as the universe expands;
 \begin{align}\label{a}
\varphi_0(t)|_{t>t_e}\simeq\left(\frac{\sqrt{c_M/(n-1)}}{\lambda}H(t)M_p^{n-3}\right)^{\frac{1}{n-2}}.
\end{align}
For simplicity, we assume potential energy of inflation is converted to its oscillation energy instantaneously at $t=t_e$. 
Once $H(t)$ crosses $H_{\rm osc}\simeq m_{\phi}/\sqrt{c_M}$, however, the mass of the AD field become positive and $\phi$ starts to oscillate around the origin. At the same time, the phase direction of the AD field $\theta$, which stays at certain direction $\theta_0$ due to the Hubble friction, receives a ``kick" from  A-term potential, so that $\phi$ starts to rotate in the complex plane. This dynamics generates the baryon asymmetry of the universe since baryon number density is represented as
  \begin{align}\label{nb}
n_{B}=-2b{\rm Im}[\phi^*\dot{\phi}]=-2b\varphi^2\dot{\theta},
\end{align}
 where $b$ is a baryon charge of $\phi$.

\subsection{Baryon asymmetry}

Let us estimate the resultant baryon asymmetry of the universe. Using E.O.M for $\phi$, the evolution of baryon number density eq.(\ref{nb}) is determined as
 \begin{align}
\dot{n_{B}}+3Hn_{B}=2b{\rm Im}\left[\phi^*\frac{\partial V}{\partial \phi^*}\right].
\end{align}
Integrating the differential equation, we get

\begin{align}
a(t)^3n_{B}(t)&=2ba_m\int^{t}_{t_e}dt'a(t')^3\lambda m_{3/2}\varphi(t')^n\sin(n\theta)\\ 
\label{n0}&\simeq2ba_m\int^{t_{\rm osc}}_{t_e}dt'a(t')^3\lambda m_{3/2}\varphi(t')^n\sin(n\theta_0).
\end{align}
where we used eq.(\ref{pot}). Since $\theta$ starts to oscillate at $t_{\rm osc}$, integration over $t>t_{\rm osc}$ also oscillates and has little contribution. To calculate eq.(\ref{n0}), usually we assume $\phi$ is always at the vacuum, i.e. $\varphi(t)=\varphi_0(t)$. Then we obtain the well-known result.
\begin{align}\label{n}
n_{B}(t_{\rm osc})&=\epsilon  m_{3/2}\varphi_0(t_{osc})^2,\\
\epsilon&=\frac{4ba_m \sqrt{c_M}\sin(n\theta_0)}{3\sqrt{n-1}\left(\frac{n-4}{n-2}+1\right)}.
\end{align}
In this paper, we point out that this conventional result can overestimate the baryon asymmetry by $\mathcal{O}(1-100)$ for the most part of the parameter region. This is because the deviation of $\varphi(t)$ from its minimum $\varphi_0(t)$ is hard to be neglected. Representing the deviation as
\begin{align}
\varphi(t)=\chi(t)\varphi_0(t),
\end{align}
we can take the effect into account and get more precise expression such as
\begin{align}
n_{B}(t_{\rm osc})&\simeq \epsilon  m_{3/2}\varphi_0(t_{osc})^2\,\overline{\chi^n},\\
\overline{\chi^n}&=\left(\frac{1}{t_{\rm osc}}\int^{t_{\rm osc}}_{t_e}\chi(t)^ndt\right).
\end{align}
In the next section, we discuss the actual value of ``efficiency factor" $\overline{\chi^n}$.

\section{Estimation of efficiency factor}

As we mentioned in the previous section, $\varphi(t)=\varphi_0(t)$ does not satisfy the E.O.M. for $t>t_e$\footnote{for $t<t_e$, the deviation $|\chi(t)-1|$ is exponentially suppressed due to the inflation.}. In fact, the E.O.M. for $\chi$ for $t>t_e$ becomes
\begin{align}\label{eom}
\chi''(z)+\frac{n-4}{n-2}\chi'(z)  =& \left[\frac{4}{9}c_M+\frac{n-3}{(n-2)^2}\right]\chi(z)\nonumber \\
& -\frac{4}{9}c_M\chi(z)^{2n-3},
\end{align}
where we take $z=\ln (t/t_e)$ as a differential variable ($(')\equiv d/dz$). We can see that there is a fixed point of $\chi$ near the unity,
\begin{align}
\chi_0=\left(1+\frac{9(n-3)}{4c_M(n-2)^2}\right)^\frac{1}{2(n-2)}.
\end{align}
However, the static solution $\chi(z)=\chi_0$ does not meet the realistic situation. As we saw in the previous section, until the end of inflation, $\varphi$ sits still at the minimum given by eq.(\ref{b}) due to the Hubble induced mass. From matching of $\varphi$ and $\varphi'$ at $z=z_e (=0)$, $\chi(z)$ must satisfy the following initial conditions:
\begin{align}\label{ini}
\chi(z_e)=c_r^\frac{1}{2(n-2)},~\chi'(z_e)=\frac{\chi(z_e)}{n-2},
\end{align}
where we define $c_r\equiv c_I/c_M$. Therefore, $\chi(t)$ has non-zero ``velocity" and inevitably oscillates around $\chi_0$. 

\begin{figure}[t]
   \centering
   \includegraphics[width=70mm]{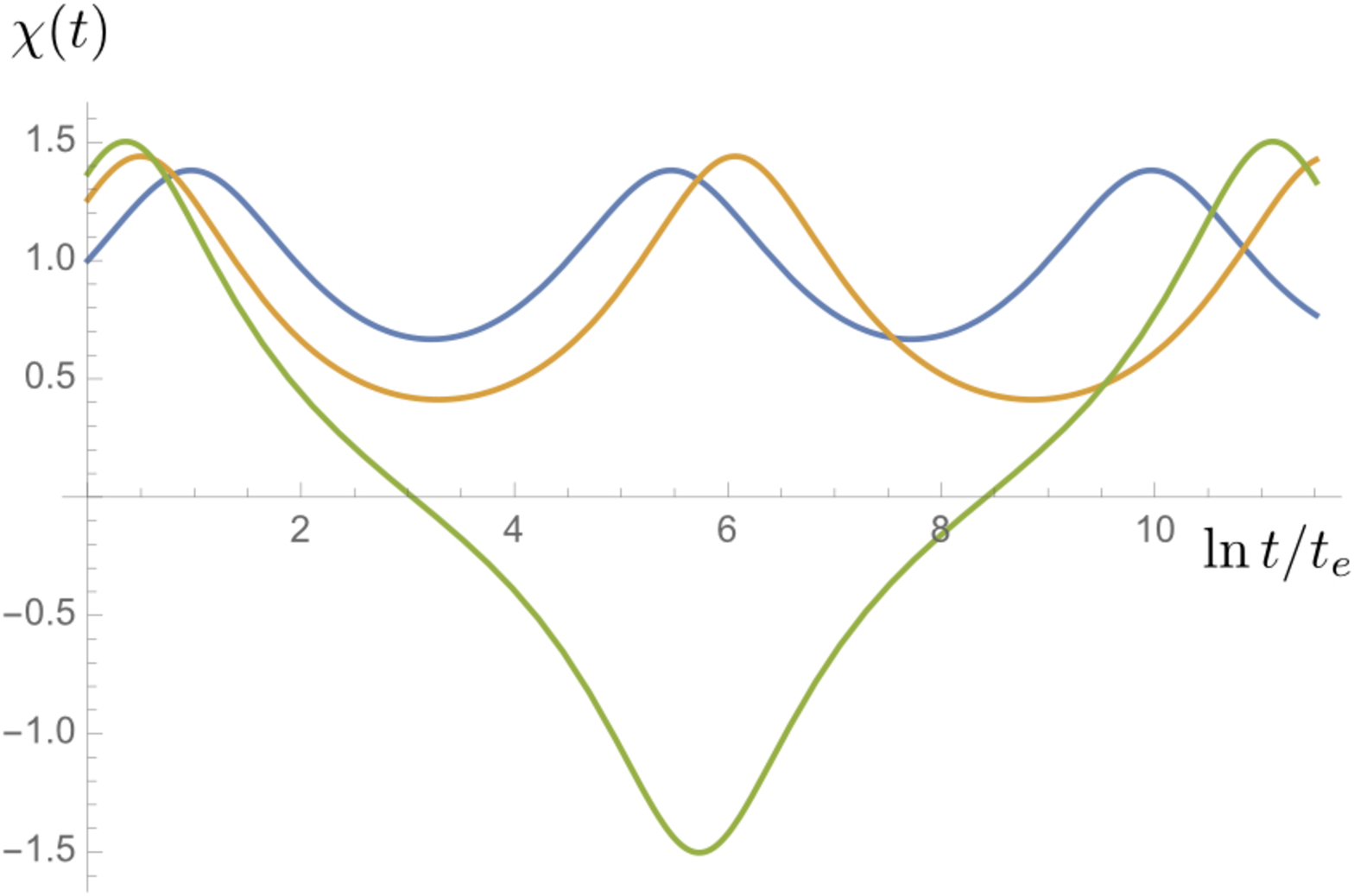}
   \caption{Evolution of $\chi$ obtained by numerically integrating eq.~(\ref{eom}) for $n=4$. Blue, yellow, green line denotes the solution for $c_r=1,2.5,3.5$ respectively.} 
   \label{fig:chi_evolution}
 \end{figure}

In particular, let us consider the case of $n=4$. For simplicity, we set $c_M=1$. As one can see from eq.~(\ref{eom}), the friction term for $\chi$ is absent for $n=4$, so that the oscillation lasts until $H(t)\sim H_{\rm osc}$. We numerically solved the dynamics of $\chi(z)$ as shown in Fig.~\ref{fig:chi_evolution}. 
 \begin{figure}[t]
   \centering
   \includegraphics[width=90mm]{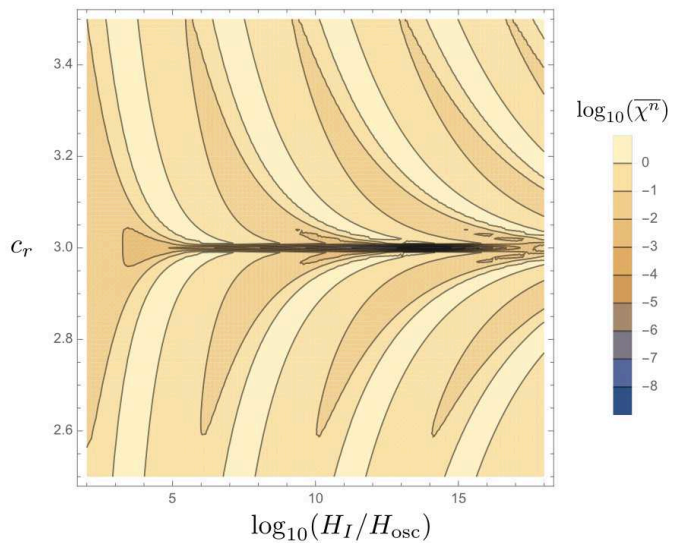}
   \caption{The contour plot of the factor $\overline{\chi^n}$ in the case of $n=4$. The vertical axis is $c_r$ and the horizontal axis is a ratio $H_I/H_{\rm osc}$} 
   \label{fig:chi_factor}
\end{figure}

Note that this oscillation dynamics has two distinct features. First, the oscillation is periodic only with respect to $z=\ln t$, i.e., $\chi(z)=\chi(z+T)$. This means that, in term of the cosmic time $t$,  the period of the oscillation increases exponentially. Therefore, it can be said that $\overline{\chi^n}$ is characterized by the value of ${\chi^n}$ just before $t_{\rm osc}$. For example, if a wave crest appears just before $t_{\rm osc}$, $\overline{\chi^n}$ would take a value of order $\chi_{\rm max}^n$. On the contrary, when a wave hollow appears, $\overline{\chi^n}$ can be a more suppressed value of order $\chi_{\rm min}^n$. Second, the branching point of the dynamics exists at $\chi(z_e)=3^{1/4}$, i.e. $c_r=3$. For $c_r<3$, $\chi$ oscillates around $\chi_0$ staying in the region $\chi>0$. On the other hand, for $c_r>3$, $\chi$ crosses over the origin and oscillates around the origin. The choice $c_r=3$ is a nothing but the unstable solution where $\chi$ approaches the origin with infinite time. Therefore, if we take $c_r\simeq3$, $\chi$ approaches to the origin closely and $\chi_{\rm min}$ can be much less than unity. 

According to these facts, we can understand the behavior of the numerical result of $\overline{\chi^n}$ shown in Fig.~\ref{fig:chi_factor}. $\overline{\chi^n}$ oscillates with respect to $H_I/H_{\rm osc}$, i.e., the duration of the $\chi$ oscillation, because of the first feature we mentioned above. Consequently, depending on the choice of $c_r$ and $H_I/H_{\rm osc}$, the baryon number density could receive accidental suppression in the $n=4$ AD baryogenesis scenario. 

We stress that this suppression mechanism is effective only when the inflation sector consists of more than two superfields. In the single superfield inflation, the value of $c_M$ and $c_I$ are not independent (see footnote \ref{f}) and the value of $c_r$ is written as $c_r=\frac{c'-1}{c'-1/2}$, which is smaller than unity for $c'>1$. Therefore we can not take $c_r\simeq3$, which is a necessary condition for the large suppression. 

\begin{figure}[t]
   \centering
   \includegraphics[width=70mm]{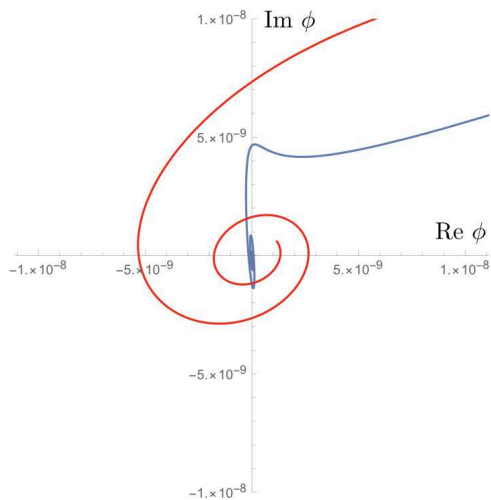}
    \caption{Dynamics of $\phi$ in the complex plane ($n=4$). The blue and red trajectories represent the cases for $c_r=2.97$ and $2.9$, respectively. It is seen that tiny difference of $\mathcal{O}(1)$ parameter $c_I$ can change the amplitude of rotation of $\phi$ in the complex plane.}
    \label{fig:dynamics_CP}

 \end{figure}

\begin{figure}[t]
   \centering
   \includegraphics[width=80mm]{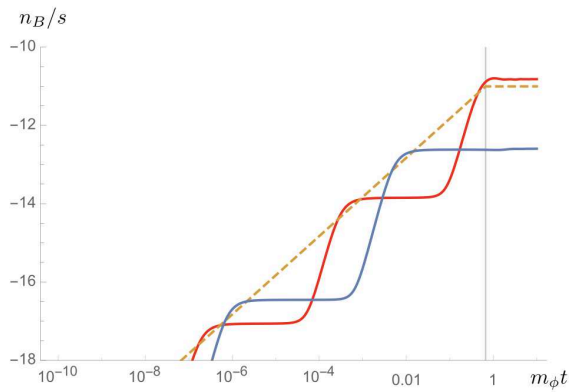}
   \caption{Evolution of the baryon number density $n_{B}$ normalized by entropy density $s$. The yellow dashed line represents the ``ideal" case where $\phi$ exactly follows the minimum ($\chi(t)\equiv1$). The blue and red lines represent actual numerical calculations for $c_r=2.97$ and $2.9$, respectively. The vertical line denotes the time when the AD field starts to oscillate around the origin. Here we take $H_I=10^{10}{\rm GeV}$, $H_{\rm osc}=10^{5}{\rm GeV}$, $T_R\sqrt{c_r}=10^8{\rm GeV}$, $\theta_0=\pi/10$, $b=1$ and $a=1$.}
   \label{fig:baryon_asym}

\end{figure}

Our discussion above is based on the semi-analytical estimation of the baryon number density eq.~(\ref{n0}). In fact, this is confirmed by the fully numerical simulation including the phase direction of the AD field $\theta$ as shown in Figs.~\ref{fig:dynamics_CP} and \ref{fig:baryon_asym}.

In the case of $n>4$ AD baryogenesis, such suppression mechanism also occurs. However, the friction term in eq.~(\ref{eom}) becomes effective and the oscillation amplitude decreases with $z=\ln t/t_e$. Therefore, $\overline{\chi^n}$ approaches to unity for $\log_{10}(H_I/H_{\rm osc})\gtrsim\mathcal{O}(1)$ and the suppression is less important. 

\section{Minimal Affleck-Dine Leptogenesis in high-scale inflation}

We have discussed the possible suppression of the baryon number density. In general, we want to avoid such a suppression in order to generate a sufficient amount of baryon asymmetry of the universe we observe today. On the other hand, the Affleck-Dine baryogenesis has a problem in high-scale inflation where the inflation scale takes a larger value $H_I\gtrsim10^{13}$GeV. During inflation the phase of the AD field obtains fluctuations $\simeq H_I/2\pi$ which result in baryonic isocurvature perturbations. In high-scale inflation the isocurvature perturbations are too large unless the field value during inflation is nearly the Planck scale, which brings difficulties to some AD baryogenesis scenarios including the minimal Affleck-Dine leptogenesis. In this section, we show that the suppression effect can make the minimal Affleck-Dine leptogenesis possible even in high-scale inflation. 

In the minimal Affleck-Dine Leptogenesis, $LH_u$ direction \cite{MURAYAMA1994349}, which possess the lepton number, is used as the AD field. This direction is lifted up by non-renormalizable term which gives mass to the neutrinos;
\begin{align}
W_{\rm AD}=&\frac{m_{\nu_i}}{2\langle H_u\rangle^2}(L_iH_u)^2\\
\equiv&\frac{\lambda}{4M_p}\Phi^4,\ \ \ {\rm for}\  \frac{\Phi^2}{2}=LH_u,
\end{align}
where $\langle H_u\rangle=(174{\rm GeV})\sin\beta$ ($\tan\beta=\langle H_u\rangle/\langle H_d\rangle$), and here we take the basis where mass matrix for neutrinos is diagonal. Therefore, $LH_u$ direction corresponds to $n=4$ AD field. Then, the lightest neutrino mass is related to $\lambda$ as
\begin{align}
m_{\nu1}=\frac{\lambda\langle H_u\rangle^2}{M_p}\simeq5.3\times10^{-9}{\rm eV}\left(\frac{\lambda}{4.2\times10^{-4}}\right).
\end{align}
Note that the value of $\lambda$ has an upper bound of order $10^{-4}$ due to the baryonic isocurvature constraint in high-scale inflation \cite{PhysRevD.90.043510,1475-7516-2008-10-017,Kawasaki2001388,PhysRevLett.83.2510,PhysRevD.62.043502}. 
When we take the inflation scale as $H_I\simeq10^{13}\rm{GeV}$, the upper bound is evaluated as $\lambda\lesssim 4.2\times10^{-4}$.
Therefore the model predicts such a very tiny neutrino mass.

In this scenario, the Affleck-Dine mechanism produces the $L$ asymmetry of the universe. Since the sphaleron process  is in thermal equilibrium, the produced $L$ asymmetry is converted to the baryon asymmetry as
\begin{align}
n_{B}\simeq -\frac{8}{23}n_{L}.
\end{align}
Consequently, the present baryon-to-entropy density is estimated as \cite{PhysRevD.42.3344}
\begin{align}\label{ba}
\frac{n_{B}}{s}
\simeq-\frac{8}{23}\frac{T_Rn_{L}(t_{\rm osc})}{4M_p^2H_{\rm osc}^2}=\epsilon\frac{8}{23}\frac{T_Rm_{3/2}}{4\sqrt{3}\lambda M_pH_{\rm osc}}.
\end{align}

Here we have to mention the finite temperature effect. In particular for the $n=4$ Affleck-Dine baryogenesis, the thermal log potential \cite{ANISIMOV2001729,PhysRevD.63.123513}
\begin{align}
V_T(\phi)\simeq c_T\alpha_s^2T^4\log\left(\frac{|\phi|^2}{T^2}\right)
\end{align}
could change the dynamics of AD fields, where $T$ is a temperature of the background plasma and $c_T=45/32$. This thermal potential behaves as a positive mass term for $\phi$, which modifies the time when the AD field starts to oscillate as
\begin{align}
H_{\rm osc}\simeq{\rm Max}[m_{\phi},~0.6\alpha_s\sqrt{\lambda}T_R].
\end{align}
We can see that in high-scale inflation, the thermal mass easily overcomes the soft mass for AD field. Consequently, we obtain the resulting baryon asymmetry by substituting $H_{\rm osc}=0.6\alpha_s\sqrt{\lambda}T_R$ into eq.(\ref{ba}) as

\begin{align}
\frac{n_{B}}{s}\simeq4.1\times10^{-11}\epsilon\left(\frac{m_{3/2}}{1{\rm TeV}}\right)\left(\frac{\lambda}{4.2\times10^{-4}}\right)^{-3/2},
\end{align}
where we assume $\alpha_s\simeq0.1$. Surprisingly, the result does not depend on the reheating temperature as long as $T_R\gtrsim m_{\phi}\lambda^{-1/2}/\alpha_s$ is satisfied \cite{PhysRevD.63.123513}. 

To realize the observed baryon asymmetry, gravitino mass $m_{3/2}$ should be related with $\lambda$ as
\begin{align}
m_{3/2}=2.1~{\rm TeV}\left(\frac{\lambda}{4.2\times10^{-4}}\right)^{3/2}\lesssim 2.1~{\rm TeV}.
\end{align}
On the other hand, unfortunately, gravitinos with such a mass decays in the era of the BBN and destroy the light elements \cite{PhysRevLett.82.4168,PhysRevD.62.023506,PhysRevD.71.083502,PhysRevD.72.043522,PhysRevD.78.065011}. To avoid the problem, gravitinos must decay before the BBN. For example, if reheating occurred via gravitational interaction, where the reheating temperature is typically $T_R\sim10^9~{\rm GeV}$, gravitino mass has a lower bound such as $m_{3/2}\gtrsim 10^4~{\rm GeV}$. Therefore, the minimal Affleck-Dine leptogenesis does not work in high-scale inflation due to the gravitino problem. One may consider the fine-tuning of the initial angle $\theta_0\ll1$ which make $\epsilon\ll1$. However, such a tiny $\theta_0$ makes the baryonic isocurvature constraint stronger and upper bound on $m_{3/2}$ become lower by $\mathcal{O}(\theta_0)$.

\begin{figure}[t]
  \centering
   \includegraphics[width=90mm]{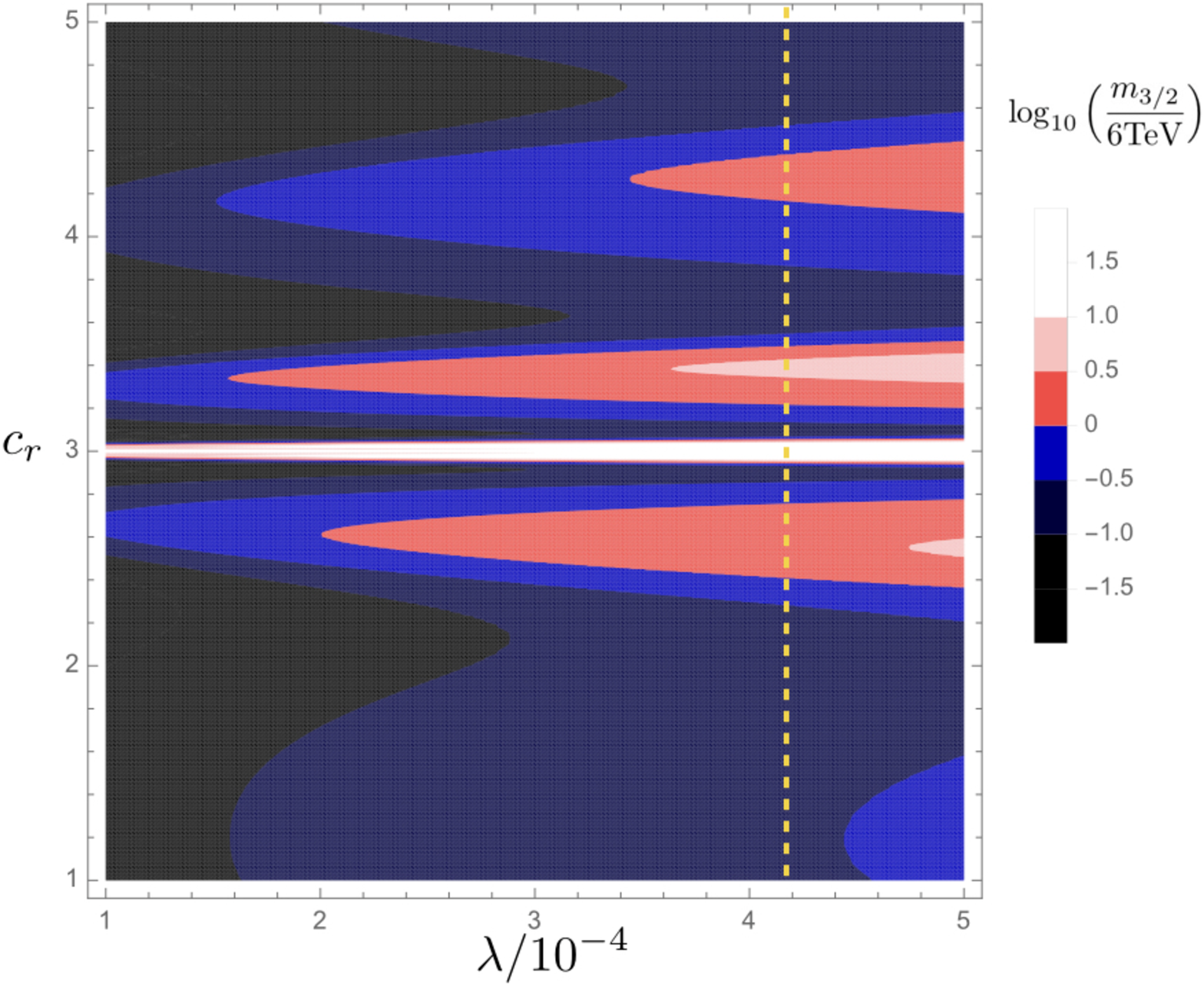}
   \includegraphics[width=90mm]{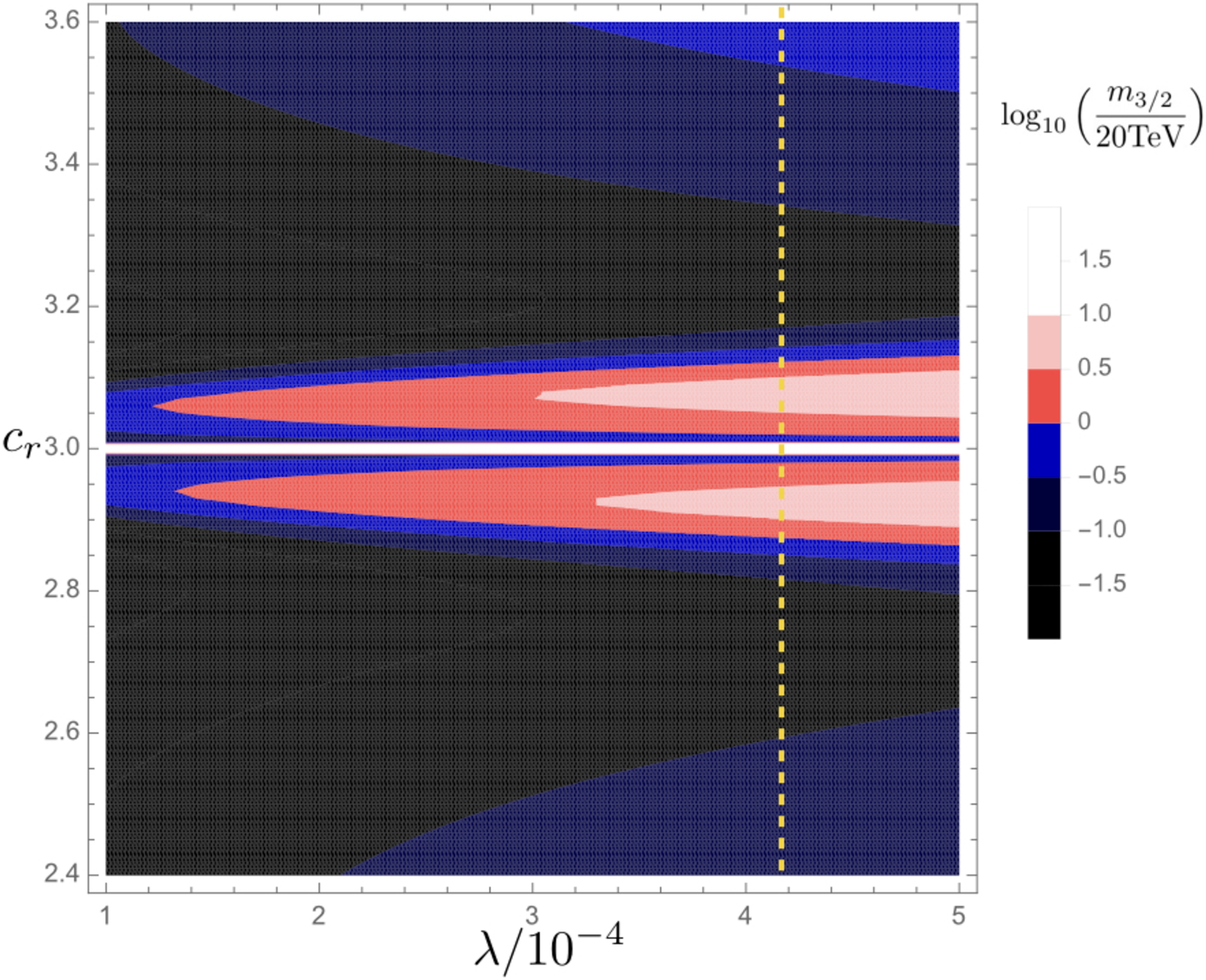}
   \caption{We plot the value of gravitino mass which consistently explains the baryon asymmetry of the universe. The blue region is excluded by the gravitino problem and red one is allowed. The up and down panels correspond to the case with $T_R=10^9$~GeV and $10^{10}~{\rm GeV}$, here we set the BBN bound on the gravitino mass as $6$~TeV and $20~{\rm TeV}$ respectively. The yellow dashed line represents the upper bound on $\lambda$ from the isocurvature perturbation.}
  \label{fig:LHu}
\end{figure}

However, it is possible for the scenario to work if we take the oscillation of the AD field into account. As we discussed in the previous sections, oscillation of the AD field leads accidental suppression of the produced baryon asymmetry as
\begin{align}
\frac{\tilde{n}_{B}}{s}\simeq\overline{\chi^4}(c_r,H_I/H_{\rm osc})\cdot\frac{{n}_{B}}{s}.
\end{align}
Consequently, the required gravitino mass to realize the observed baryon number density becomes heavier as
\begin{align}\label{ub}
m_{3/2}\simeq2.1~{\rm TeV}\left(\overline{\chi^4}(c_r,H_I/H_{\rm osc})\right)^{-1}.
\end{align}
We numerically calculate the gravitino mass which makes $n_{B}/s\simeq8.7\times10^{-11}$ \cite{1674-1137-38-9-090001} and plot on the $(c_r,~\lambda)$ - plane in Fig.~\ref{fig:LHu}.
From the figure, it is seen that we can take heavier gravitino masses and hence evade the BBN constraint.

\section{Summary and disscussions}

In this paper, we have performed the more precise estimation of the produced baryon asymmetry in the $n=4$ AD baryogenesis. Due to the fact that the AD condensate oscillates around its minimum after inflation, the efficiency of the generation of the baryon asymmetry decreases depending on the choice of dimension-less $\mathcal{O}(1)$ parameter $c_r$ and the quantity $H_I/H_{\rm osc}$. We found that this suppression mechanism make the minimal AD leptogenesis scenario viable even in high-scale inflation. 

In our analysis, we assume inflation suddenly ends and switches to the matter domination era. Although this simplification does not change the suppression mechanism, dependence of $\overline{\chi^4}$ on $c_I$ and $H_I/H_\text{osc}$ could be slightly modified. We note that the choice of $c_M$, which we set unity in the paper, also does not change the result.  

Finally, let us comment on the evolution of the fluctuation $\delta\phi$. When $\phi$ approaches the origin, the effective mass of $\delta\phi$ becomes negative due to the negative Hubble induced mass. Therefore tachyonic resonance \cite{PhysRevLett.87.011601} would take place and the fluctuations $\delta\phi$ grow exponentially. However, the resonance is not effective because the oscillation time scale ($=$ period) of the AD field is an order of the Hubble time and hence the AD field oscillates only several times before producing baryon asymmetry. After the soft SUSY breaking mass dominates the dynamics, the fluctuations of the AD field generally grow and form non-topological solitons called Q-balls~\cite{Kusenko199846,KUSENKO1997108,Enqvist1998309,COLEMAN1985263,PhysRevD.64.123515,PhysRevD.62.023512,ENQVIST1999321}. Large Q-balls may decay after the electroweak phase transition, which makes lepton-baryon number conversion difficult since the sphaleron process is ineffective. In the case of the $LH_u$ direction, however, the existence of the supersymmetric $\mu$ term can prevent the AD field from forming Q-balls. 

\section*{Acknowledgement}
F.H. would like to thank Jeong-Pyong Hong and Yutaro Shoji for helpful comments. This work is supported by MEXT KAKENHI Grant Number 15H05889 (M. K.), JSPS KAKENHI Grant Number 17K05434 (M. K.) and also by the World Premier International Research Center Initiative (WPI), MEXT, Japan. F. H. is supported by JSPS Research Fellowship for Young Scientists Grant Number 17J07391.

\bibliography{ADBG}

\begin{thebibliography}{29}%
\makeatletter
\providecommand \@ifxundefined [1]{%
 \@ifx{#1\undefined}
}%
\providecommand \@ifnum [1]{%
 \ifnum #1\expandafter \@firstoftwo
 \else \expandafter \@secondoftwo
 \fi
}%
\providecommand \@ifx [1]{%
 \ifx #1\expandafter \@firstoftwo
 \else \expandafter \@secondoftwo
 \fi
}%
\providecommand \natexlab [1]{#1}%
\providecommand \enquote  [1]{``#1''}%
\providecommand \bibnamefont  [1]{#1}%
\providecommand \bibfnamefont [1]{#1}%
\providecommand \citenamefont [1]{#1}%
\providecommand \href@noop [0]{\@secondoftwo}%
\providecommand \href [0]{\begingroup \@sanitize@url \@href}%
\providecommand \@href[1]{\@@startlink{#1}\@@href}%
\providecommand \@@href[1]{\endgroup#1\@@endlink}%
\providecommand \@sanitize@url [0]{\catcode `\\12\catcode `\$12\catcode
  `\&12\catcode `\#12\catcode `\^12\catcode `\_12\catcode `\%12\relax}%
\providecommand \@@startlink[1]{}%
\providecommand \@@endlink[0]{}%
\providecommand \url  [0]{\begingroup\@sanitize@url \@url }%
\providecommand \@url [1]{\endgroup\@href {#1}{\urlprefix }}%
\providecommand \urlprefix  [0]{URL }%
\providecommand \Eprint [0]{\href }%
\providecommand \doibase [0]{http://dx.doi.org/}%
\providecommand \selectlanguage [0]{\@gobble}%
\providecommand \bibinfo  [0]{\@secondoftwo}%
\providecommand \bibfield  [0]{\@secondoftwo}%
\providecommand \translation [1]{[#1]}%
\providecommand \BibitemOpen [0]{}%
\providecommand \bibitemStop [0]{}%
\providecommand \bibitemNoStop [0]{.\EOS\space}%
\providecommand \EOS [0]{\spacefactor3000\relax}%
\providecommand \BibitemShut  [1]{\csname bibitem#1\endcsname}%
\let\auto@bib@innerbib\@empty
\bibitem [{\citenamefont {Affleck}\ and\ \citenamefont
  {Dine}(1985)}]{AFFLECK1985361}%
  \BibitemOpen
  \bibfield  {author} {\bibinfo {author} {\bibfnamefont {I.}~\bibnamefont
  {Affleck}}\ and\ \bibinfo {author} {\bibfnamefont {M.}~\bibnamefont {Dine}},\
  }\href {\doibase http://dx.doi.org/10.1016/0550-3213(85)90021-5} {\bibfield
  {journal} {\bibinfo  {journal} {Nuclear Physics B}\ }\textbf {\bibinfo
  {volume} {249}},\ \bibinfo {pages} {361 } (\bibinfo {year}
  {1985})}\BibitemShut {NoStop}%
\bibitem [{\citenamefont {Dine}\ \emph {et~al.}(1996)\citenamefont {Dine},
  \citenamefont {Randall},\ and\ \citenamefont {Thomas}}]{DINE1996291}%
  \BibitemOpen
  \bibfield  {author} {\bibinfo {author} {\bibfnamefont {M.}~\bibnamefont
  {Dine}}, \bibinfo {author} {\bibfnamefont {L.}~\bibnamefont {Randall}}, \
  and\ \bibinfo {author} {\bibfnamefont {S.}~\bibnamefont {Thomas}},\ }\href
  {\doibase http://dx.doi.org/10.1016/0550-3213(95)00538-2} {\bibfield
  {journal} {\bibinfo  {journal} {Nuclear Physics B}\ }\textbf {\bibinfo
  {volume} {458}},\ \bibinfo {pages} {291 } (\bibinfo {year}
  {1996})}\BibitemShut {NoStop}%
\bibitem [{\citenamefont {Mukaida}\ and\ \citenamefont
  {Nakayama}(2013)}]{1475-7516-2013-01-017}%
  \BibitemOpen
  \bibfield  {author} {\bibinfo {author} {\bibfnamefont {K.}~\bibnamefont
  {Mukaida}}\ and\ \bibinfo {author} {\bibfnamefont {K.}~\bibnamefont
  {Nakayama}},\ }\href {http://stacks.iop.org/1475-7516/2013/i=01/a=017}
  {\bibfield  {journal} {\bibinfo  {journal} {JCAP}\ }\textbf {\bibinfo
  {volume} {2013}},\ \bibinfo {pages} {017} (\bibinfo {year}
  {2013})}\BibitemShut {NoStop}%
\bibitem [{\citenamefont {Fukugita}\ and\ \citenamefont
  {Yanagida}(1986)}]{FUKUGITA198645}%
  \BibitemOpen
  \bibfield  {author} {\bibinfo {author} {\bibfnamefont {M.}~\bibnamefont
  {Fukugita}}\ and\ \bibinfo {author} {\bibfnamefont {T.}~\bibnamefont
  {Yanagida}},\ }\href {\doibase
  http://dx.doi.org/10.1016/0370-2693(86)91126-3} {\bibfield  {journal}
  {\bibinfo  {journal} {Physics Letters B}\ }\textbf {\bibinfo {volume}
  {174}},\ \bibinfo {pages} {45 } (\bibinfo {year} {1986})}\BibitemShut
  {NoStop}%
\bibitem [{\citenamefont {Kuzmin}\ \emph {et~al.}(1985)\citenamefont {Kuzmin},
  \citenamefont {Rubakov},\ and\ \citenamefont {Shaposhnikov}}]{KUZMIN198536}%
  \BibitemOpen
  \bibfield  {author} {\bibinfo {author} {\bibfnamefont {V.}~\bibnamefont
  {Kuzmin}}, \bibinfo {author} {\bibfnamefont {V.}~\bibnamefont {Rubakov}}, \
  and\ \bibinfo {author} {\bibfnamefont {M.}~\bibnamefont {Shaposhnikov}},\
  }\href {\doibase http://dx.doi.org/10.1016/0370-2693(85)91028-7} {\bibfield
  {journal} {\bibinfo  {journal} {Physics Letters B}\ }\textbf {\bibinfo
  {volume} {155}},\ \bibinfo {pages} {36 } (\bibinfo {year}
  {1985})}\BibitemShut {NoStop}%
\bibitem [{\citenamefont {Murayama}\ and\ \citenamefont
  {Yanagida}(1994)}]{MURAYAMA1994349}%
  \BibitemOpen
  \bibfield  {author} {\bibinfo {author} {\bibfnamefont {H.}~\bibnamefont
  {Murayama}}\ and\ \bibinfo {author} {\bibfnamefont {T.}~\bibnamefont
  {Yanagida}},\ }\href {\doibase
  http://dx.doi.org/10.1016/0370-2693(94)91164-9} {\bibfield  {journal}
  {\bibinfo  {journal} {Physics Letters B}\ }\textbf {\bibinfo {volume}
  {322}},\ \bibinfo {pages} {349 } (\bibinfo {year} {1994})}\BibitemShut
  {NoStop}%
\bibitem [{\citenamefont {Kasuya}\ and\ \citenamefont
  {Kawasaki}(2006)}]{PhysRevD.74.063507}%
  \BibitemOpen
  \bibfield  {author} {\bibinfo {author} {\bibfnamefont {S.}~\bibnamefont
  {Kasuya}}\ and\ \bibinfo {author} {\bibfnamefont {M.}~\bibnamefont
  {Kawasaki}},\ }\href {\doibase 10.1103/PhysRevD.74.063507} {\bibfield
  {journal} {\bibinfo  {journal} {Phys. Rev. D}\ }\textbf {\bibinfo {volume}
  {74}},\ \bibinfo {pages} {063507} (\bibinfo {year} {2006})}\BibitemShut
  {NoStop}%
\bibitem [{\citenamefont {Harigaya}\ \emph {et~al.}(2014)\citenamefont
  {Harigaya}, \citenamefont {Kamada}, \citenamefont {Kawasaki}, \citenamefont
  {Mukaida},\ and\ \citenamefont {Yamada}}]{PhysRevD.90.043510}%
  \BibitemOpen
  \bibfield  {author} {\bibinfo {author} {\bibfnamefont {K.}~\bibnamefont
  {Harigaya}}, \bibinfo {author} {\bibfnamefont {A.}~\bibnamefont {Kamada}},
  \bibinfo {author} {\bibfnamefont {M.}~\bibnamefont {Kawasaki}}, \bibinfo
  {author} {\bibfnamefont {K.}~\bibnamefont {Mukaida}}, \ and\ \bibinfo
  {author} {\bibfnamefont {M.}~\bibnamefont {Yamada}},\ }\href {\doibase
  10.1103/PhysRevD.90.043510} {\bibfield  {journal} {\bibinfo  {journal} {Phys.
  Rev. D}\ }\textbf {\bibinfo {volume} {90}},\ \bibinfo {pages} {043510}
  (\bibinfo {year} {2014})}\BibitemShut {NoStop}%
\bibitem [{\citenamefont {Kasuya}\ \emph {et~al.}(2008)\citenamefont {Kasuya},
  \citenamefont {Kawasaki},\ and\ \citenamefont
  {Takahashi}}]{1475-7516-2008-10-017}%
  \BibitemOpen
  \bibfield  {author} {\bibinfo {author} {\bibfnamefont {S.}~\bibnamefont
  {Kasuya}}, \bibinfo {author} {\bibfnamefont {M.}~\bibnamefont {Kawasaki}}, \
  and\ \bibinfo {author} {\bibfnamefont {F.}~\bibnamefont {Takahashi}},\ }\href
  {http://stacks.iop.org/1475-7516/2008/i=10/a=017} {\bibfield  {journal}
  {\bibinfo  {journal} {JCAP}\ }\textbf {\bibinfo {volume} {2008}},\ \bibinfo
  {pages} {017} (\bibinfo {year} {2008})}\BibitemShut {NoStop}%
\bibitem [{\citenamefont {Kawasaki}\ and\ \citenamefont
  {Takahashi}(2001)}]{Kawasaki2001388}%
  \BibitemOpen
  \bibfield  {author} {\bibinfo {author} {\bibfnamefont {M.}~\bibnamefont
  {Kawasaki}}\ and\ \bibinfo {author} {\bibfnamefont {F.}~\bibnamefont
  {Takahashi}},\ }\href {\doibase
  https://doi.org/10.1016/S0370-2693(01)00957-1} {\bibfield  {journal}
  {\bibinfo  {journal} {Physics Letters B}\ }\textbf {\bibinfo {volume}
  {516}},\ \bibinfo {pages} {388 } (\bibinfo {year} {2001})}\BibitemShut
  {NoStop}%
\bibitem [{\citenamefont {Enqvist}\ and\ \citenamefont
  {McDonald}(1999{\natexlab{a}})}]{PhysRevLett.83.2510}%
  \BibitemOpen
  \bibfield  {author} {\bibinfo {author} {\bibfnamefont {K.}~\bibnamefont
  {Enqvist}}\ and\ \bibinfo {author} {\bibfnamefont {J.}~\bibnamefont
  {McDonald}},\ }\href {\doibase 10.1103/PhysRevLett.83.2510} {\bibfield
  {journal} {\bibinfo  {journal} {Phys. Rev. Lett.}\ }\textbf {\bibinfo
  {volume} {83}},\ \bibinfo {pages} {2510} (\bibinfo {year}
  {1999}{\natexlab{a}})}\BibitemShut {NoStop}%
\bibitem [{\citenamefont {Enqvist}\ and\ \citenamefont
  {McDonald}(2000)}]{PhysRevD.62.043502}%
  \BibitemOpen
  \bibfield  {author} {\bibinfo {author} {\bibfnamefont {K.}~\bibnamefont
  {Enqvist}}\ and\ \bibinfo {author} {\bibfnamefont {J.}~\bibnamefont
  {McDonald}},\ }\href {\doibase 10.1103/PhysRevD.62.043502} {\bibfield
  {journal} {\bibinfo  {journal} {Phys. Rev. D}\ }\textbf {\bibinfo {volume}
  {62}},\ \bibinfo {pages} {043502} (\bibinfo {year} {2000})}\BibitemShut
  {NoStop}%
\bibitem [{\citenamefont {Harvey}\ and\ \citenamefont
  {Turner}(1990)}]{PhysRevD.42.3344}%
  \BibitemOpen
  \bibfield  {author} {\bibinfo {author} {\bibfnamefont {J.~A.}\ \bibnamefont
  {Harvey}}\ and\ \bibinfo {author} {\bibfnamefont {M.~S.}\ \bibnamefont
  {Turner}},\ }\href {\doibase 10.1103/PhysRevD.42.3344} {\bibfield  {journal}
  {\bibinfo  {journal} {Phys. Rev. D}\ }\textbf {\bibinfo {volume} {42}},\
  \bibinfo {pages} {3344} (\bibinfo {year} {1990})}\BibitemShut {NoStop}%
\bibitem [{\citenamefont {Anisimov}\ and\ \citenamefont
  {Dine}(2001)}]{ANISIMOV2001729}%
  \BibitemOpen
  \bibfield  {author} {\bibinfo {author} {\bibfnamefont {A.}~\bibnamefont
  {Anisimov}}\ and\ \bibinfo {author} {\bibfnamefont {M.}~\bibnamefont
  {Dine}},\ }\href {\doibase http://dx.doi.org/10.1016/S0550-3213(01)00550-8}
  {\bibfield  {journal} {\bibinfo  {journal} {Nuclear Physics B}\ }\textbf
  {\bibinfo {volume} {619}},\ \bibinfo {pages} {729 } (\bibinfo {year}
  {2001})}\BibitemShut {NoStop}%
\bibitem [{\citenamefont {Fujii}\ \emph {et~al.}(2001)\citenamefont {Fujii},
  \citenamefont {Hamaguchi},\ and\ \citenamefont
  {Yanagida}}]{PhysRevD.63.123513}%
  \BibitemOpen
  \bibfield  {author} {\bibinfo {author} {\bibfnamefont {M.}~\bibnamefont
  {Fujii}}, \bibinfo {author} {\bibfnamefont {K.}~\bibnamefont {Hamaguchi}}, \
  and\ \bibinfo {author} {\bibfnamefont {T.}~\bibnamefont {Yanagida}},\ }\href
  {\doibase 10.1103/PhysRevD.63.123513} {\bibfield  {journal} {\bibinfo
  {journal} {Phys. Rev. D}\ }\textbf {\bibinfo {volume} {63}},\ \bibinfo
  {pages} {123513} (\bibinfo {year} {2001})}\BibitemShut {NoStop}%
\bibitem [{\citenamefont {Kawasaki}\ \emph {et~al.}(1999)\citenamefont
  {Kawasaki}, \citenamefont {Kohri},\ and\ \citenamefont
  {Sugiyama}}]{PhysRevLett.82.4168}%
  \BibitemOpen
  \bibfield  {author} {\bibinfo {author} {\bibfnamefont {M.}~\bibnamefont
  {Kawasaki}}, \bibinfo {author} {\bibfnamefont {K.}~\bibnamefont {Kohri}}, \
  and\ \bibinfo {author} {\bibfnamefont {N.}~\bibnamefont {Sugiyama}},\ }\href
  {\doibase 10.1103/PhysRevLett.82.4168} {\bibfield  {journal} {\bibinfo
  {journal} {Phys. Rev. Lett.}\ }\textbf {\bibinfo {volume} {82}},\ \bibinfo
  {pages} {4168} (\bibinfo {year} {1999})}\BibitemShut {NoStop}%
\bibitem [{\citenamefont {Kawasaki}\ \emph {et~al.}(2000)\citenamefont
  {Kawasaki}, \citenamefont {Kohri},\ and\ \citenamefont
  {Sugiyama}}]{PhysRevD.62.023506}%
  \BibitemOpen
  \bibfield  {author} {\bibinfo {author} {\bibfnamefont {M.}~\bibnamefont
  {Kawasaki}}, \bibinfo {author} {\bibfnamefont {K.}~\bibnamefont {Kohri}}, \
  and\ \bibinfo {author} {\bibfnamefont {N.}~\bibnamefont {Sugiyama}},\ }\href
  {\doibase 10.1103/PhysRevD.62.023506} {\bibfield  {journal} {\bibinfo
  {journal} {Phys. Rev. D}\ }\textbf {\bibinfo {volume} {62}},\ \bibinfo
  {pages} {023506} (\bibinfo {year} {2000})}\BibitemShut {NoStop}%
\bibitem [{\citenamefont {Kawasaki}\ \emph {et~al.}(2005)\citenamefont
  {Kawasaki}, \citenamefont {Kohri},\ and\ \citenamefont
  {Moroi}}]{PhysRevD.71.083502}%
  \BibitemOpen
  \bibfield  {author} {\bibinfo {author} {\bibfnamefont {M.}~\bibnamefont
  {Kawasaki}}, \bibinfo {author} {\bibfnamefont {K.}~\bibnamefont {Kohri}}, \
  and\ \bibinfo {author} {\bibfnamefont {T.}~\bibnamefont {Moroi}},\ }\href
  {\doibase 10.1103/PhysRevD.71.083502} {\bibfield  {journal} {\bibinfo
  {journal} {Phys. Rev. D}\ }\textbf {\bibinfo {volume} {71}},\ \bibinfo
  {pages} {083502} (\bibinfo {year} {2005})}\BibitemShut {NoStop}%
\bibitem [{\citenamefont {Ichikawa}\ \emph {et~al.}(2005)\citenamefont
  {Ichikawa}, \citenamefont {Kawasaki},\ and\ \citenamefont
  {Takahashi}}]{PhysRevD.72.043522}%
  \BibitemOpen
  \bibfield  {author} {\bibinfo {author} {\bibfnamefont {K.}~\bibnamefont
  {Ichikawa}}, \bibinfo {author} {\bibfnamefont {M.}~\bibnamefont {Kawasaki}},
  \ and\ \bibinfo {author} {\bibfnamefont {F.}~\bibnamefont {Takahashi}},\
  }\href {\doibase 10.1103/PhysRevD.72.043522} {\bibfield  {journal} {\bibinfo
  {journal} {Phys. Rev. D}\ }\textbf {\bibinfo {volume} {72}},\ \bibinfo
  {pages} {043522} (\bibinfo {year} {2005})}\BibitemShut {NoStop}%
\bibitem [{\citenamefont {Kawasaki}\ \emph {et~al.}(2008)\citenamefont
  {Kawasaki}, \citenamefont {Kohri}, \citenamefont {Moroi},\ and\ \citenamefont
  {Yotsuyanagi}}]{PhysRevD.78.065011}%
  \BibitemOpen
  \bibfield  {author} {\bibinfo {author} {\bibfnamefont {M.}~\bibnamefont
  {Kawasaki}}, \bibinfo {author} {\bibfnamefont {K.}~\bibnamefont {Kohri}},
  \bibinfo {author} {\bibfnamefont {T.}~\bibnamefont {Moroi}}, \ and\ \bibinfo
  {author} {\bibfnamefont {A.}~\bibnamefont {Yotsuyanagi}},\ }\href {\doibase
  10.1103/PhysRevD.78.065011} {\bibfield  {journal} {\bibinfo  {journal} {Phys.
  Rev. D}\ }\textbf {\bibinfo {volume} {78}},\ \bibinfo {pages} {065011}
  (\bibinfo {year} {2008})}\BibitemShut {NoStop}%
\bibitem [{\citenamefont {Olive}\ and\ \citenamefont
  {Group}(2014)}]{1674-1137-38-9-090001}%
  \BibitemOpen
  \bibfield  {author} {\bibinfo {author} {\bibfnamefont {K.}~\bibnamefont
  {Olive}}\ and\ \bibinfo {author} {\bibfnamefont {P.~D.}\ \bibnamefont
  {Group}},\ }\href {http://stacks.iop.org/1674-1137/38/i=9/a=090001}
  {\bibfield  {journal} {\bibinfo  {journal} {Chinese Physics C}\ }\textbf
  {\bibinfo {volume} {38}},\ \bibinfo {pages} {090001} (\bibinfo {year}
  {2014})}\BibitemShut {NoStop}%
\bibitem [{\citenamefont {Felder}\ \emph {et~al.}(2001)\citenamefont {Felder},
  \citenamefont {Garc\'{\i}a-Bellido}, \citenamefont {Greene}, \citenamefont
  {Kofman}, \citenamefont {Linde},\ and\ \citenamefont
  {Tkachev}}]{PhysRevLett.87.011601}%
  \BibitemOpen
  \bibfield  {author} {\bibinfo {author} {\bibfnamefont {G.}~\bibnamefont
  {Felder}}, \bibinfo {author} {\bibfnamefont {J.}~\bibnamefont
  {Garc\'{\i}a-Bellido}}, \bibinfo {author} {\bibfnamefont {P.~B.}\
  \bibnamefont {Greene}}, \bibinfo {author} {\bibfnamefont {L.}~\bibnamefont
  {Kofman}}, \bibinfo {author} {\bibfnamefont {A.}~\bibnamefont {Linde}}, \
  and\ \bibinfo {author} {\bibfnamefont {I.}~\bibnamefont {Tkachev}},\ }\href
  {\doibase 10.1103/PhysRevLett.87.011601} {\bibfield  {journal} {\bibinfo
  {journal} {Phys. Rev. Lett.}\ }\textbf {\bibinfo {volume} {87}},\ \bibinfo
  {pages} {011601} (\bibinfo {year} {2001})}\BibitemShut {NoStop}%
\bibitem [{\citenamefont {Kusenko}\ and\ \citenamefont
  {Shaposhnikov}(1998)}]{Kusenko199846}%
  \BibitemOpen
  \bibfield  {author} {\bibinfo {author} {\bibfnamefont {A.}~\bibnamefont
  {Kusenko}}\ and\ \bibinfo {author} {\bibfnamefont {M.}~\bibnamefont
  {Shaposhnikov}},\ }\href {\doibase
  https://doi.org/10.1016/S0370-2693(97)01375-0} {\bibfield  {journal}
  {\bibinfo  {journal} {Physics Letters B}\ }\textbf {\bibinfo {volume}
  {418}},\ \bibinfo {pages} {46 } (\bibinfo {year} {1998})}\BibitemShut
  {NoStop}%
\bibitem [{\citenamefont {Kusenko}(1997)}]{KUSENKO1997108}%
  \BibitemOpen
  \bibfield  {author} {\bibinfo {author} {\bibfnamefont {A.}~\bibnamefont
  {Kusenko}},\ }\href {\doibase
  http://dx.doi.org/10.1016/S0370-2693(97)00584-4} {\bibfield  {journal}
  {\bibinfo  {journal} {Physics Letters B}\ }\textbf {\bibinfo {volume}
  {405}},\ \bibinfo {pages} {108 } (\bibinfo {year} {1997})}\BibitemShut
  {NoStop}%
\bibitem [{\citenamefont {Enqvist}\ and\ \citenamefont
  {McDonald}(1998)}]{Enqvist1998309}%
  \BibitemOpen
  \bibfield  {author} {\bibinfo {author} {\bibfnamefont {K.}~\bibnamefont
  {Enqvist}}\ and\ \bibinfo {author} {\bibfnamefont {J.}~\bibnamefont
  {McDonald}},\ }\href {\doibase https://doi.org/10.1016/S0370-2693(98)00271-8}
  {\bibfield  {journal} {\bibinfo  {journal} {Physics Letters B}\ }\textbf
  {\bibinfo {volume} {425}},\ \bibinfo {pages} {309 } (\bibinfo {year}
  {1998})}\BibitemShut {NoStop}%
\bibitem [{\citenamefont {Coleman}(1985)}]{COLEMAN1985263}%
  \BibitemOpen
  \bibfield  {author} {\bibinfo {author} {\bibfnamefont {S.}~\bibnamefont
  {Coleman}},\ }\href {\doibase http://dx.doi.org/10.1016/0550-3213(85)90286-X}
  {\bibfield  {journal} {\bibinfo  {journal} {Nuclear Physics B}\ }\textbf
  {\bibinfo {volume} {262}},\ \bibinfo {pages} {263 } (\bibinfo {year}
  {1985})}\BibitemShut {NoStop}%
\bibitem [{\citenamefont {Kasuya}\ and\ \citenamefont
  {Kawasaki}(2001)}]{PhysRevD.64.123515}%
  \BibitemOpen
  \bibfield  {author} {\bibinfo {author} {\bibfnamefont {S.}~\bibnamefont
  {Kasuya}}\ and\ \bibinfo {author} {\bibfnamefont {M.}~\bibnamefont
  {Kawasaki}},\ }\href {\doibase 10.1103/PhysRevD.64.123515} {\bibfield
  {journal} {\bibinfo  {journal} {Phys. Rev. D}\ }\textbf {\bibinfo {volume}
  {64}},\ \bibinfo {pages} {123515} (\bibinfo {year} {2001})}\BibitemShut
  {NoStop}%
\bibitem [{\citenamefont {Kasuya}\ and\ \citenamefont
  {Kawasaki}(2000)}]{PhysRevD.62.023512}%
  \BibitemOpen
  \bibfield  {author} {\bibinfo {author} {\bibfnamefont {S.}~\bibnamefont
  {Kasuya}}\ and\ \bibinfo {author} {\bibfnamefont {M.}~\bibnamefont
  {Kawasaki}},\ }\href {\doibase 10.1103/PhysRevD.62.023512} {\bibfield
  {journal} {\bibinfo  {journal} {Phys. Rev. D}\ }\textbf {\bibinfo {volume}
  {62}},\ \bibinfo {pages} {023512} (\bibinfo {year} {2000})}\BibitemShut
  {NoStop}%
\bibitem [{\citenamefont {Enqvist}\ and\ \citenamefont
  {McDonald}(1999{\natexlab{b}})}]{ENQVIST1999321}%
  \BibitemOpen
  \bibfield  {author} {\bibinfo {author} {\bibfnamefont {K.}~\bibnamefont
  {Enqvist}}\ and\ \bibinfo {author} {\bibfnamefont {J.}~\bibnamefont
  {McDonald}},\ }\href {\doibase
  http://dx.doi.org/10.1016/S0550-3213(98)00695-6} {\bibfield  {journal}
  {\bibinfo  {journal} {Nuclear Physics B}\ }\textbf {\bibinfo {volume}
  {538}},\ \bibinfo {pages} {321 } (\bibinfo {year}
  {1999}{\natexlab{b}})}\BibitemShut {NoStop}%
\end{thebibliography}%

\end{document}